\documentclass[a4paper,aps,noeprint,twocolumn,secnumarabic,
  balancelastpage,amsmath,amssymb,nofootinbib,longbibliography,
  preprint,10pt
]{revtex4-1}
\usepackage{pifont}
\usepackage{graphicx}
\usepackage{dcolumn}
\usepackage{subfigure}
\usepackage{bm}
\usepackage[colorlinks=true,breaklinks=true,allcolors=blue]{hyperref}
\usepackage{url}
\usepackage{xcolor}
\usepackage{enumitem}
 
\makeatletter

\makeatletter
\let\cat@comma@active\@empty
\makeatother

\AtBeginDocument{%
    \newwrite\bibnotes
    \def\bibnotesext{Notes.bib}
    \immediate\openout\bibnotes=\jobname\bibnotesext
    \immediate\write\bibnotes{@CONTROL{REVTEX41Control}}
    \immediate\write\bibnotes{@CONTROL{%
    apsrev41Control,author="08",editor="1",pages="1",title="0",year="1"}}
     \if@filesw
     \immediate\write\@auxout{\string\citation{apsrev41Control}}%
    \fi
}%

\begin{document}

\title{Quantum adiabatic cycles and their breakdown}

\author{Nicol\`o Defenu}
\email{ndefenu@phys.ethz.ch}
\affiliation{Institut f\"ur Theoretische Physik, ETH Z\"urich, Wolfgang-Pauli-Str.\,27 Z\"urich, Switzerland}

\begin{abstract}
The assumption that quasi-static transformations do not quantitatively alter the equilibrium expectation of observables is at the heart of thermodynamics and, in the quantum realm, its validity may be confirmed by the application of adiabatic perturbation theory. Yet, this scenario does not straightforwardly apply to Bosonic systems whose excitation energy is slowly driven through the zero. Here, we prove that the universal slow dynamics of such systems is always non-adiabatic and the quantum corrections to the equilibrium observables become rate independent for any dynamical protocol in the slow drive limit. These findings overturn the common expectation for quasi-static processes as they demonstrate that a system as simple and general as the quantum harmonic oscillator, does not allow for a slow-drive limit, but it always displays sudden quench dynamics.
\end{abstract}

\maketitle
\section{Introduction}
 Quasi-static processes are thermodynamic transformations which happen slow enough not to cause any sizeable variation to the instantaneous equilibrium solution of the problem\,\cite{landau2013statistical}. A convenient mathematical representation for these processes considers a system, initially at equilibrium, whose Hamiltonian is slowly varied in time $H(\delta\cdot t)$ with a rate much smaller than any internal scale of the system. Under proper assumptions on the analyticity of the evolution and of  the thermodynamic functions, an analytic scaling $\sim\delta^{2}$ for the dynamical corrections to the equilibrium expectations may be predicted\,\cite{zwerger2008limited}. 

In the quantum realm, the concept of ``adiabaticity", i.e. the possibility to realise an equilibrium state by a quasi-static process, is crucial to quantum computation, where non-trivial correlations in the system ground state are generated by a slow variation of the Hamiltonian parameters\,\cite{farhi2001quantum}. The possibility of such manipulation is granted by the quantum adiabatic theorem\,\cite{born1928beweis,kato1950adiabatic,avron1999adiabatic}, which ensures that the outcome of the adiabatic procedure will converge to the ground-state of the final Hamiltonian in the $\delta\to 0$ limit. 

The prototypical model for quantum adiabatic dynamics is the Landau-Zener (LZ) problem, which describes the excitation probability of a two level system ramped over an avoided eigenvalue crossing\,\cite{zener1932non,landau1965quantum}. In analogy with the classical case, the exact solution of the LZ problem features dynamical corrections which vanish exponentially in the slow drive limit. However, at a quantum critical point (QCP) an actual eigenvalue crossing appears\,\cite{sachdev1999quantum} and non-analytic corrections $\sim\delta^{\theta}$ to the adiabatic observables emerge, according to the Kibble-Zurek mechanism (KZM), where the $\theta$-scaling only depends on the equilibrium critical exponents\,\cite{zurek1985cosmological,zurek2005dynamics}. Interestingly, an exact description of KZM in thermodynamic systems with purely Fermionic quasi-particles can be obtained by relating the quasi-particle dynamics to an infinite number of LZ transitions with momentum dependent minimal gaps\,\cite{dziarmaga2005dynamics, dziarmaga2010dynamics}. Therefore, the LZ problem has remained up to now one of the most precious tools to understand defects formation in quantum systems\,\cite{damski2005simplest}. 

Nevertheless, several quantum many-body systems feature strongly interacting QCPs  and no quadratic effective field theory in terms of Fermi quasi-particles can be constructed. The validity of KZM scaling in these systems can be shown by adiabatic perturbation theory, which, under proper scaling assumptions, is able to reproduce the expected non-analytic scaling for the defect density $n_{\rm exc}\approx\delta^{\theta}$\,\cite{polkovnikov2005universal}. Notice that the assumptions made in Ref.\,\cite{polkovnikov2005universal} in order to derive the KZM prediction for generic quantum many-body systems may not apply to systems with  competing interactions\,\cite{defenu2019universal}.
	
 More in general, the adiabatic perturbation theory approach cannot be applied to harmonic systems with Bosonic quasi-particles as the perturbative assumption is violated by Bose statistics, which allows macroscopic population in the excited resonant states\,\cite{degrandi2009adiabatic, galtbayar2020solvable}. 
Moreover, several critical systems ranging from quantum magnets and cavity systems to superfluids and supersolids can be effectively described by harmonic Bose quasi-particles, whose excitation energy gradually vanishes approaching the QCP\,\cite{sachdev1999quantum,zwerger2008limited}.

\begin{figure}[ht]
\parbox{0.98\linewidth}{\includegraphics[width=\linewidth]{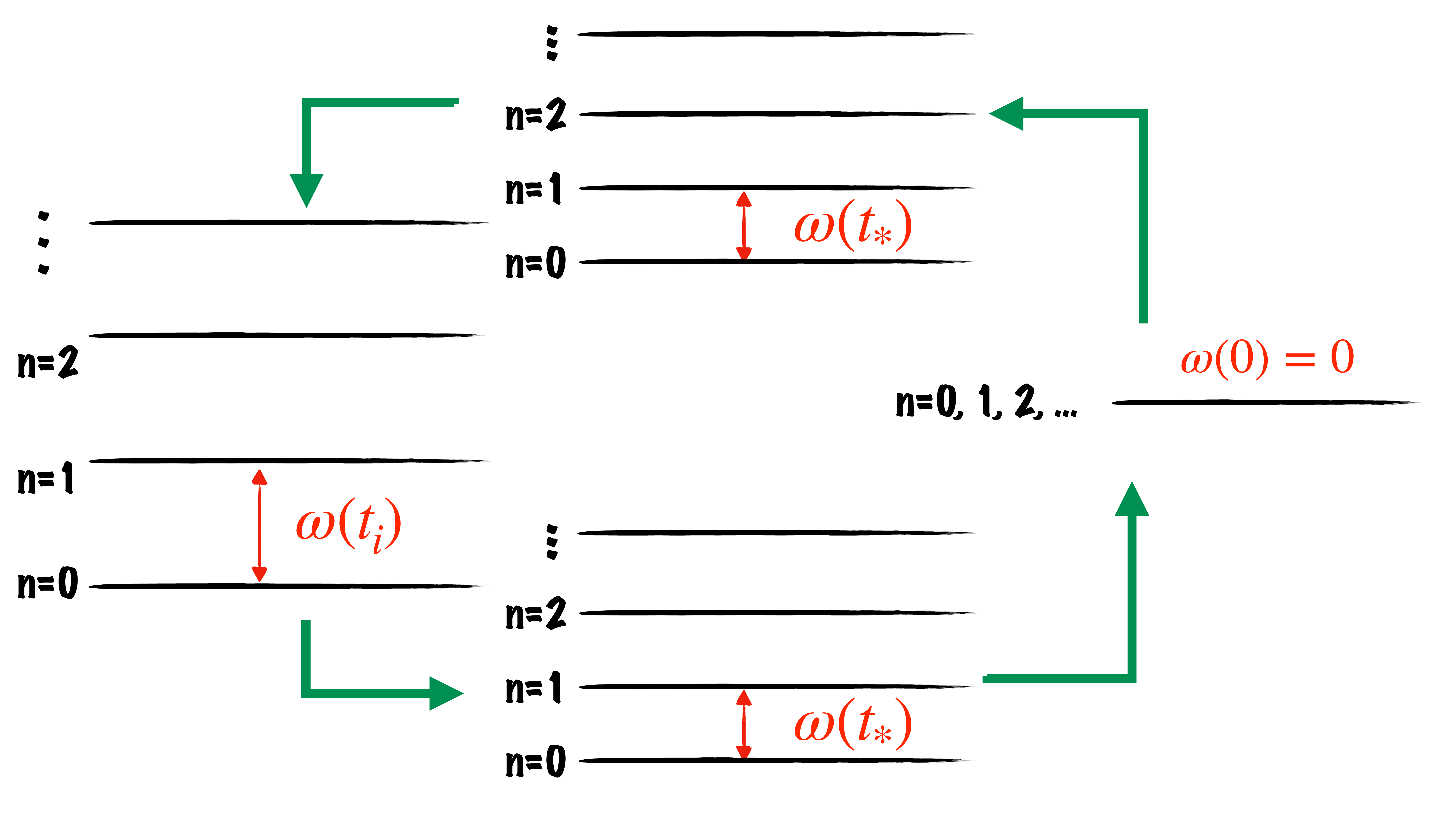}}
\caption{\label{Fig1} Schematic representation of the quantum adiabatic cycle under study. The system is prepared in the ground state of the Hamiltonian ($n=0$) with a regular, well separated, spectrum at the initial time $-t_{i}$. Each excited state is labeled by a growing integer $n$. Then, the Hamiltonian is dynamically driven in such a way to reduce the spectral gap of the system $\omega(t_{*})\ll \omega(-t_{i})$ (i.e. followong the lower green arrows), until the instantaneous spectrum becomes fully degenerate $\omega(t=0)\simeq 0$ (on the right in the picture). Finally, the drive protocol is inverted and the initial Hamiltonian is restored (following the upper green arrows).}\end{figure}

In the following, we investigate quantum adiabatic cycles across a  QCP, where infinite many excitation levels become degenerate (corresponding to the case of Bose statistics for the excitations), see Fig.\,\ref{Fig1}. The general assumptions of the quantum adiabatic theorem do not hold in this case and no-general result over the dynamical corrections to the adiabatic observables is known\,\cite{born1928beweis,kato1950adiabatic,avron1999adiabatic}. We prove that adiabaticity breakdown is a universal feature of these systems independently of the considered drive rate and shape. These results justify and extend recent studies concerning non-adiabatic defect formation $n_{\rm exc}\approx O(1)$ in fully-connected many-body systems and in single-mode harmonic Hamiltonians with  analytic $\sim t^{2}$ drives\,\cite{bachmann2017dynamical, defenu2018dynamical}.

One of the fundamental consequences of these findings concern the full characterisation of defect formation in critical quantum many-body systems, as we provide the missing piece of information to summarise universal adiabatic dynamics as follows:
\begin{itemize}
\item Finite systems: $n_{\rm exc}\approx\delta^{2}$.
\item Interacting QCPs: $n_{\rm exc}\approx\delta^{\theta}$.
\item Harmonic Bose quasi-particles: $n_{\rm exc}\approx O(1)$.
\end{itemize}
The first class is conveniently represented by the LZ model, while the second one can be treated by adiabatic perturbation theory. The present investigations focus on the third class, where the dynamical corrections are always non-adiabatic, i.e. rate independent, but for which no general result was known up to now. 

It is worth noting that the aforementioned regimes for defect scaling may also appear in a given quantum system depending on the type of dynamical protocol performed, see the results section. In particular, for a system with harmonic Bose quasi-particles, the non-analytic $\delta^{\theta}$ scaling may be found for dynamical protocols terminating exactly at the QCP (regime 1). While any actual crossing of the gapless point will lead to a finite defect density $n_{\rm exc}\approx O(1)$ (regime 2). Therefore, dynamical quasi-static transformations of Bosonic systems across QCPs are the main focus of the present paper.  

Before proceeding further with the analysis, it is convenient to discuss the aforementioned picture in the context of the existing literature. Seminal studies on the Kibble-Zurek scaling across QCPs have been performed in Refs.\,\cite{zurek2005dynamics,polkovnikov2005universal,dziarmaga2005dynamics,damski2005simplest} in the context of many-body systems with Fermi quasiparticles. The extension of these analyses to the case of Bose modes, such as spin-waves, has been limited to the case of quenches in the vicinity of a critical point\,\cite{polkovnikov2008breakdown,degrandi2009adiabatic}, where regime (1) has been analysed only for linear scaling of the square frequency $\omega(t)^{2}\approx \delta\cdot t$. Also, Refs.\,\cite{polkovnikov2008breakdown,degrandi2009adiabatic} consider a continuum ensemble of non-interacting Bose quasi-particles with gapless spectrum rather than a single mode. Then, the non-adiabatic phase observed in Refs.\,\cite{polkovnikov2008breakdown,degrandi2009adiabatic} is not the consequence of the crossing of the critical point (which is not discussed there), but of the infra-red divergence of spin-wave contributions in low-dimensions, which also causes the disappearance of continuous symmetry breaking transitions in $d\leq 2$, according to the Mermin-Wagner theorem\,\cite{mermin1966absence,frohlich1976phase,codello2015critical}.

First mathematical evidences of the existence of regime (2) have been found in Ref.\,\cite{bachmann2017dynamical}, where the scaling of the single mode  gap was assumed to be linear ($\omega(t)^{2}\approx t^{2}$). This solution is more straightforward due to the homogeneous scaling of the time parameter and the position operator $\omega(t)x^{2}\propto (t\,x)^{2}$. In the physics context, these results have been used to justify the anomalous defect scaling numerically observed in the LMG model\,\cite{acevedo2014new,defenu2018dynamical}. 

In this work we are going to prove that the existence of regime (2) is actually a generic feature of any dynamical protocol, crossing a QCP with pure bosonic quasi-particles. The amount of heat and the number of defects generated at the end of these dynamical manipulations will be shown to be universal functions, which do not depend on the drive rate nor on the peculiar drive shape, but only on the leading scaling exponent in the time-dependent frequency expansion $\omega(t)\approx (\delta|t|)^{z\nu}+\cdots$.  Moreover, our analysis will extend the observations of Refs.\,\cite{polkovnikov2008breakdown,degrandi2009adiabatic} for dynamical evolutions terminating in the vicinity of the QCP to any scaling exponent $z\nu$.

\section{Results}
\label{sec_res}
In order to prove our picture, let us consider a single dynamically driven Harmonic mode with Hamiltonian
\begin{align}
\label{hoh}
H(t)=\frac{1}{2}\left(p^{2}+\omega(t)^{2}x^{2}\right).
\end{align}
A part from its fundamental interest, the Hamiltonian in Eq.\,\eqref{hoh} faithfully describes the quantum fluctuations of many-body systems with fully-connected cavity mediated interactions such as the Dicke\,\cite{dicke1954coherence} or the Lipkin-Meshkov-Glick (LMG) models\,\cite{lipkin1965validity,meshkov1965validity,glick1965validity,dusuel2004finite,vidal2006finite,defenu2018dynamical} and, more in general, models which feature a collective single mode excitation, such as the BCS model\,\cite{dusuel2005finite}.

The dynamics described by Eq.\,\eqref{hoh} cannot be explicitly solved in general, but an explicit solution can be obtained for the scaling form
\begin{equation}
\label{freq_scal}
\omega(t)^{2}=\left(\delta |t|\right)^{2z\nu}\end{equation}
where $\delta>0$ is the drive rate and the exponent $z\nu>0$ represents the  gap scaling exponent. In the following we are going to show that any time-dependent shape $\omega(t)$, which crosses the QCP at $t=0$, can be reduced to the form in Eq.\,\eqref{freq_scal} in the $\delta\to 0$ limit.

Eq.\,\eqref{hoh} with the time-dependent frequency in Eq.\,\eqref{freq_scal} may be regarded as effectively describing a many-body system ramped across its QCP, in the spirit of Refs.\,\cite{polkovnikov2008breakdown,bachmann2017dynamical,defenu2018dynamical}. Within this perspective the exponent $z\nu$ represents the dynamical critical exponent for the gap scaling\,\cite{sachdev1999quantum}. However, it is worth noting that in the framework of the effective theory in Eq.\,\eqref{hoh} the quantity $z\nu$ in Eq.\,\eqref{freq_scal} is merely a tuneable parameter describing the dynamical protocol and it is not directly related to any critical behaviour displayed by the effective model at equilibrium.  

As long as the the spectral gap remains finite at all instants ($\omega(t)\neq 0\,\, \forall\,\, t$) the scaling of the defect density and the corrections to the dynamical observables with respect to the instantaneous equilibrium expectation can be predicted using adiabatic perturbation theory\,\cite{polkovnikov2008breakdown,degrandi2009adiabatic}, see also Chap.\,3 of Ref.\,\cite{lincoln2010understanding}. In addition, as anticipated in the introduction, two universal regimes are observed according to the scaling of the observables in the quasi-static limit $\delta\to 0$:
\begin{enumerate}
\item Kibble-Zurek regime (half cycle).
\item Universal non-adiabaticity (full cycle).
\end{enumerate}
Regime (1) occurs for a half-cycle $t\in [-t_{i},0]$ (with $t_{i}\propto 1/\delta$) and features non-analytic corrections to the adiabatic expectations appearing at $t = 0$ (where $\omega(0)=0$). Such corrections cannot be captured by the standard perturbative approach, but can be predicted by the KZM scaling argument.  On the contrary for a full cycle $t\in [-t_{i},t_{i}]$ the critical point is actually crossed and the system enters in the non-adiabatic regime, where the leading correction to the observables expectation does not depend on $\delta$.  We refer to this latter scenario as regime (2). It should be stressed that the notation $z\nu$ for the frequency scaling in Eq.\,\eqref{freq_scal} is employed in order to make contact with the traditional Kibble-Zurek picture in many-body systems, but in our case the exponent $z\nu$ is just an effective quantity, which is not connected with the equilibrium critical scaling of any specific model.

The picture outlined above naturally follows from the solution of the model under study. The dynamical eigen-functions of the Hamiltonian in Eq.\,\eqref{hoh} can be written in terms of a single time dependent parameter: the effective width $\xi(t)$, see the definition in Supplementary Eq.\,(2). Then, the dynamics of the quantum problem may be obtained by the solution of the Ermakov-Milne equation, which describes the evolution of the effective width $\xi(t)$, see Supplementary Eq.\,(5).

 First, it is convenient to rewrite the Hamiltonian in Eq.\,\eqref{hoh} as a rate independent one by introducing the transformations 
\begin{align}
\label{dim_trans}
t = \delta^{-\frac{z\nu}{1+z\nu}}s,\qquad x= \delta^{-\frac{z\nu}{2+2z\nu}}\tilde{x}
\end{align}
which reduce the dynamics of the model in Eq.\,\eqref{hoh} to the $\delta=1$ case, see the Supplementary Methods 2. The expressions for the fidelity and defect density of the model, given in Supplementary Eqs.\,(12) and (13),  are invariant under the transformations in Eq.\,\eqref{dim_trans} in such a way that the fidelity and excitation density at real times can be obtained by $\tilde{\xi}(s)=\lim_{\delta\to 1}\xi(t)$ and $\tilde{\Omega}(s)^{2}=s^{2z\nu}$, provided that the endpoint of the dynamics is rescaled accordingly.

\subsection{Regime 1 (Kibble-Zurek scaling)} 
An adiabatic cycle is realised when the system starts in the instantaneous ground state of the equilibrium Hamiltonian, i.e. $\displaystyle{\lim_{t \to -t_{i}}\psi(t)=\psi_{0}^{\rm ad}(-t_{i})}$, where $\psi_{0}^{\rm ad}$ is the adiabatic state obtained replacing the constant frequency with the time-dependent one in the equilibrium ground-state\,\cite{dabrowski2016time}. Accordingly, one has to impose the boundary conditions
\begin{align}
\label{bound_cond}
\lim_{t\to-t_{i}}\xi(t)^{2}=\frac{1}{2\omega(t)};\quad \lim_{t\to-t_{i}}\dot{\xi}(t)^{2}=0.
\end{align} Following the exact solution given in the Supplementary Methods 3, the time-dependent width $\xi(t)$ and its time derivative attain a finite value in the $t\to 0$ limit. However, a finite result for the width $\xi(t)$ at  $\omega(t)=0$ corresponds to a vanishing fidelity, $f(t)\propto\sqrt{\omega(t)}$, see the Supplementary Eq. (13). Consequently, the defect density diverges as $n_{\rm exc}(t)\propto1/\omega(t)$, see Supplementary Eq. (12), but the heat (or excess energy) remains finite 
 \begin{align}
 \label{heat}
 \lim_{t\to 0}Q(t)\simeq \lim_{t\to 0}\omega(t)n_{exc}(t)\propto \delta^{\frac{z\nu}{1+z\nu}},
\end{align}
where $n_{\rm exc}$ represents the excitation density and the power-law scaling $\theta=z\nu/(1+z\nu)$ perfectly reproduces the celebrated Kibble-Zurek result\,\cite{dziarmaga2010dynamics}.

The result in Eq.\,\eqref{heat} may be also obtained by the impulse-adiabatic approximation at the basis of the KZM result\,\cite{degrandi2009adiabatic,dziarmaga2010dynamics} . Indeed, as long as the instantaneous gap remains large with respect to the drive rate $\dot{\omega}(t)\ll\omega(t)^{2}$ the dynamical state may be safely approximated by the adiabatic one $\psi_{0}(t)\simeq \psi_{0}^{\rm ad}(t)$. This approximation breaks down at the freezing time $t_{*}$ such that the adiabatic condition is violated $\dot{\omega}(t_{ *})\simeq\omega(t_{*})^{2}$. For $t>t_{*}$ the system enters in the impulse regime and the state remains frozen at $\psi_{0}^{\rm ad}(t_{*})$ with frequency $\omega(t_{*})\propto \delta^{z\nu/(1+z\nu)}$ all the way down to $t=0$. Then, the excess energy at the endpoint of the dynamics reads
\begin{align}
\label{heat_kzm}
\int_{-\infty}^{+\infty}\psi_{0}^{\rm ad,*}(t_{*})H(0)\psi_{0}^{\rm ad}(t_{*})\approx \delta^{\frac{z\nu}{1+z\nu}}
\end{align}
which reproduces the exact result in Eq.\,\eqref{heat} as well as the traditional Kibble-Zurek picture for many body systems\,\cite{zurek1996cosmological, delcampo2014universality}. The result in Eq.\,\eqref{heat_kzm} provides a first evidence of the validity of the model Hamiltonian in Eq.\,\eqref{hoh} as an effective tool to represent many-body critical dynamics.
\begin{figure*}[t!]
\centering
\includegraphics[width=1.\textwidth]{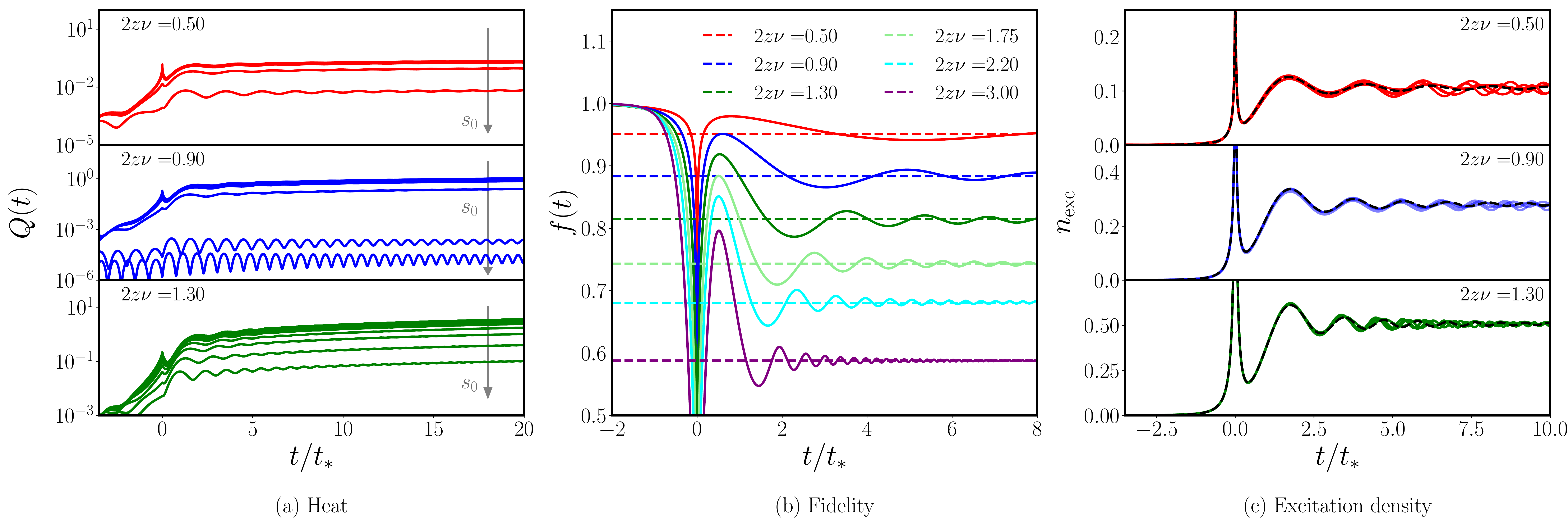}
\caption{\label{Fig2} 
Characterisation of quantum adiabatic cycles. (a)  Heat generated during a gapped cycle with time dependent gap $\omega(t)^{2}=\left(t_{0}+\delta\,|t|\right)^{2z\nu}$, with minimal gap $\omega_{0}=t_{0}^{2z\nu}$, drive rate $\delta$ and scaling exponent $z\nu$. The curves are shown as a function of the time $t$ in units of $t_{*}\propto \delta^{-z\nu/(1+z\nu)}$. Notice that the minimal gap $\omega_{0}$ is reached at the time $t_{0}$. Each sub-panel reports various curves for increasing values of  $s_{0}=\delta^{-1/(1+z\nu)}\omega_{0}^{1/z\nu}$. The values of $s_{0}\in[0,10]$ grow in the direction of the arrow. The generated heat vanishes in the large $s_{0}$ limit.  Panel (b): the fidelity of the model for different values of $z\nu=\{0.5,0.9,1.3,1.75,2.2,3\}$ (solid lines from top to bottom) is compared with the asymptotic result in Eq.\,\eqref{asy_fid} (dashed horizontal lines).  Panel (c): the numerical results for the number of excitations, defined by Eq.\,(12) in the Supplementary Methods 1, have been calculated  within the generalised dynamical model described by Eq.\,\eqref{gen_dyn_prot}. Different values of $\gamma\in[0,0.01]$, which quantify the extent of the non-universal correction see Eq.\,\eqref{gen_dyn_prot}, produce different curves  for $t>t_{*}$ (solid lines). However, the long-time limit converges to the same asymptotic value. The exact analytic solutions at $\gamma=0$ are shown as black dashed lines.}
\end{figure*}

\subsection{Regime 2 (universal non-adiabaticity)} A full cycle is realised when the system actually crosses the QCP at $t=0$. There, the driving protocol in Eq.\,\eqref{freq_scal} is non-analytic, but a proper solution can be achieved requiring that the dynamical state and its time derivative remain continuous at all times. Thus, defining the quantities
$\displaystyle{\lim_{t \to 0^{\pm}}}\xi(t)=\xi_{\pm}$, a proper continuity condition for the time dependent width reads
\begin{align}
\label{bound_cond2}
\xi_{+} =\xi_{-} ,\quad
\dot{\xi}_{+} =\dot{\xi}_{-} .
\end{align}
For a gapped cycle where $\displaystyle{\lim_{t\to 0}\omega(t)\neq 0}$ in the $\delta\to 0$ limit, the  conditions in Eq.\,\eqref{bound_cond2} is automatically satisfied and the $\xi_{+}(t)$ solution at $t>0$ approaches the same form as in the first branch $t\leq0$ of the dynamics, as required by the quantum adiabatic theorem, see Fig.\,\ref{Fig2} and the Supplementary Methods 3 C. Then, a gapped cycle always remains adiabatic and the corrections to scaling can be described within the same adiabatic perturbation theory picture developed in Refs.\,\cite{polkovnikov2008breakdown,degrandi2009adiabatic} for the $z\nu=1/2$ case.

For a gapless cycle, represented by Eq.\,\eqref{freq_scal} with  $t\in [-t_{i},t_{i}]$,  the quasi-static limit ($\delta\to 0$)  becomes rate independent, yielding the fidelity end excitations density expressions
\begin{align}
\label{asy_exc_den}
\lim_{t\to\infty} n_{\rm exc}(t)&=\tan\left(\frac{\pi}{2+2z\nu}\right)^{-2}\\
\label{asy_fid}
\lim_{t\to\infty} f(t)&=\sin\left(\frac{\pi}{2+2z\nu}\right)
\end{align}
as detailed in the Supplementary Methods 3 B.  The expressions in Eqs.\,\eqref{asy_exc_den} and\,\eqref{asy_fid} are universal with respect to rate variations, as it was already evidenced in the peculiar $z\nu=1$ case by Ref.\,\cite{bachmann2017dynamical}, where asymptotic analysis  yielded $f(\infty)=1/\sqrt{2}\quad\forall\,\,\delta$ in agreement with the result in Eq.\,\eqref{asy_fid}.

In addition, the result in Eq.\,\eqref{asy_exc_den} remains finite for any finite $z\nu$ and it only quadratically vanishes as $z\nu$ approaches zero, proving that the non-adiabatic phase does not depend on the choice of the drive scaling, but it is rather a general feature of Bosonic quantum systems. Interestingly, in the $z\nu\to\infty$ limit the system reaches what could be called an ``anti-adiabatic" phase, where the ground state fidelity completely vanishes at the end of the cycle. The approach between the numerical solution for a finite ramp extension (solid lines) 
and the exact asymptotic expressions in Eq.\,\eqref{asy_fid} and\,\eqref{asy_exc_den} is shown in Fig.\,\ref{Fig2}.
\subsection{Universality} 
Albeit the absence of any proper scaling behaviour, the results in Eqs.\,\eqref{asy_exc_den} and\,\eqref{asy_fid} are as much universal as the traditional KZM result, in the sense that they exactly describe the slow drive limit $\delta\to 0$ of any dynamical protocol which crosses the critical point. Indeed, given a general time dependent control parameter $\lambda(\delta\,t)$ the dynamics close to the critical point can be expanded  according to
\begin{align}
\label{gen_dyn_prot}
\omega'(t)^{2}\simeq (\delta t)^{2z\nu}+\gamma'(\delta t)^{n}+\cdots
\end{align}
where the integer exponent $n\in \mathbb{N}$ represents any analytic correction to critical scaling (but the same argument will apply to a non-analytic one, as long as it remains irrelevant in the $t\to 0$ limit, i.e. $2z\nu<n$). Then, applying the transformation in Eq.\,\eqref{dim_trans} one obtains the result  $\omega'(s)^{2}\simeq s^{2z\nu}+\gamma s^{n}$ where $\gamma=\delta^{(n-2z\nu)/(1+z\nu)}\gamma'$, which vanishes in the $\delta\to 0$ and reproduces the effective model considered here. 

Moreover, we have numerically verified that our analytic solution accurately describes any drive $\omega'(t)$ such that $|\omega'(t_{*})-\omega(t_{*})|^{2}\ll \omega(t_{*})^{2}$, as it is shown in Fig.\,\ref{Fig2}. There, the numerical integration of the Supplementary Eq.\,(5) with the frequency in Eq.\,\eqref{gen_dyn_prot} and different $\gamma$ values (solid curves) is compared with the analytic result for the dynamical protocol in Eq\,\eqref{freq_scal} (black dashed lines). The resulting curves for the number of excitations with different $\gamma$ values only differ in the oscillations at large times $t\gg t_{*}$, but these oscillatory terms are irrelevant  as they are washed away in the $t\to\infty$ limit.

\section{Discussion} 

The aforementioned picture for the dynamics of the Harmonic model does not only describe the simple Hamiltonian in Eq.\,\eqref{hoh}, but it also applies to conformal invariant systems confined by a time-dependent harmonic potential, such as the Calogero model\,\cite{haas1996hamiltonian}, the 1-dimensional Tonks girardeu\,\cite{minguzzi2005exact}, the trapped 2D Bose gas\,\cite{pitaevskii1997breathing}, the unitary 3D Fermi gas\,\cite{castin2012unitary} and the 2D Fermi gas, far from its crossover regime\,\cite{murthy2019quantum}. The Ermakov-Milne equation that regulates the dynamics of the model in Eq.\,\eqref{hoh} has been also used to study defect formation in a cosmological context\,\cite{matacz1994coherent,carvalho2004scalar,dabrowski2016time}, see Ref.\,\cite{perelomov1986generalized} for an overview. Moreover, a generalisation of the Ermakov-Milne equation is obtained in all dimensions by the variational treatment of the Gross-Pitaevskii equation\,\cite{perez1997dynamics}. 

More in general, our description of the Kibble-Zurek mechanism can be applied to any many-body system, whose dynamics may be approximated by an ensemble of harmonic spin-waves according to the time-dependent Hartee-Fock approximation\,\cite{hartree1928wave,fock1930,bogolyubov1947izvestiya}. In the Supplementary Note 1 an account of this procedure is given for $O(N)$ symmetric models with long-range couplings in 1-dimension, where the Hartee-Fock method becomes exact in the large-$N$ limit\,\cite{moshe2003quantum,berges2007quantum,chandran2013equilibration} (the so-called spherical model\,\cite{vojta1996quantum,defenu2017criticality}). In the last few decades, $O(N)$ field theories constituted the testbed for most calculations in critical phenomena\,\cite{wilson1974renormalization, brezin1976renormalization,brezin1976renormalization,efrati2014real,kleinert2001critical,codello2013o(n), codello2015critical} and are, even currently, a continuous source of novel universal phenomena\,\cite{fei2014critical, yabunaka2017surprises, defenu2020fate,connelly2020universal}. Our analysis shows that the universal picture derived in the present work for the Hamiltonian in Eq.\,\eqref{hoh} describes the scaling of the fidelity and the defect density in large-$N$ $O(N)$ models in the strong long-range regime, see the Supplementary Note 1 . 

Thanks to the Harmonic nature of the Hamiltonian in Eq.\,\eqref{hoh} we have been able to derive a comprehensive picture for defects formation across the $\omega(t)=0$ quantum critical point, where infinitely many excitation levels become degenerate. The present solution proves that the dynamical crossing of an infinitely degenerate quantum critical point is  non-adiabatic independently on the smallness of the rate $\delta$ and on the functional form of the drive $\omega(t)$. Adiabaticity is only recovered for a sub-power law scaling of the drive, i.e. in the $z\nu\to0$ limit. In contrast, any dynamics terminating in the vicinity of the fully degenerate critical point yields power law corrections, which can be described by the celebrated Kibble-Zurek mechanism.

The Kibble-Zurek scaling is traditionally derived within the adiabatic-impulse approximation discussed below Eq.\,\eqref{heat} and may be also justified in the more rigorous framework of adiabatic perturbation theory\,\cite{polkovnikov2005universal}. Both these descriptions fail in regime (2) of the harmonic oscillator dynamics due to the infinite number of excited states collapsing at $\omega(t)=0$. Indeed, the impulse adiabatic approximation assumes that the dynamical state is frozen at $|\psi^{\rm ad}_{0}(t_{*})\rangle$ in the entire range $t\in [-t_{*},t_{*}]$ of the dynamics. The dynamical correction to the energy at any instant of time ($t>-t_{*}$) derives from the overlap between such state and the hierarchy of adiabatic excited states $c_{n0}(t)\approx\langle\psi^{\rm ad}_{n}(t)|\psi^{\rm ad}_{0}(t_{*})\rangle$\,\cite{dabrowski2016time}. Then, the probability distribution for the excitation number $n$ remains fixed at the instant $-t_{*}$ in the entire inner regime of the dynamics, so that defects generated at $t>-t_{*}$ are effectively discarded.

In the conventional case, where a critical point with finite degeneracy is crossed, the impulse-adiabatic approximation is justified since most of the defects generated in range $-t_{*}<t<0$ annihilate at the opposite side of the cycle $0<t<t_{*}$ so that the defect distribution can be approximated by the  one at $t=-t_{*}$\,\cite{dziarmaga2010dynamics}. However, this is not the case of an infinitely degenerate quantum critical point, where the exact dynamical state also has a finite overlap with high-energy states at large-$n$. The tunnelling between such states and the adiabatic ground-state is suppressed due to the large energy separation, forbidding defects recombination. As a result, a finite fraction of the wave-function density is dispersed in the high energy portion of the spectrum after crossing the QCP and the unit fidelity cannot be recovered for any $t>0$.

The possibility to manipulate a quantum system in its ground-state heavily relies on the adiabatic properties of quasi-static transformations and it is crucial to quantum technology applications\,\cite{farhi2001quantum,lechner2015quantum}. Yet, the quantum adiabatic theorem only applies to dynamical systems with finite ground-state degeneracy\,\cite{born1928beweis,kato1950adiabatic,avron1999adiabatic}, while for the infinite degenerate case no general result was known up to now. In principle, one could have expected that a particular dynamical protocol could be devised to achieve a proper quasi-static transformation also for  quantum system dynamically driven across infinite degenerate quantum critical points. 

This is actually not the case, as we have proven that any dynamical protocol which reduces the excitation energy of an harmonic Hamiltonian down to the zero always produces a non-adiabatic outcome. Indeed, the excitation density and the fidelity results at the end of a general quasi-static transformation are universal and only depend on the drive shape, but not on its quench rate $\delta$ as long as a full cycle across the QCP is performed, see Eqs.\,\eqref{asy_exc_den} and\,\eqref{asy_fid}. This is not the case for driving protocols terminating in the vicinity of the QCP, as they remain adiabatic, see the result in Eq.\,\eqref{heat} and Refs.\,\cite{polkovnikov2008breakdown,degrandi2009adiabatic}. The present analysis unveils that a universal description of quasi-static processes can be also achieved outside the traditional assumptions of the quantum adiabatic theorem, opening to the possibility that adiabaticity breakdown is a universal feature of QCPs with infinite state degeneracy also beyond the harmonic result discussed here.

\acknowledgements I acknowledge fruitful discussions with  T.~Enss, G.\,M.~Graf, M.~Kastner and G.~Morigi on this problem. I also thank T.~Enss, G.~Gori, G.\,M. Graf and A.~Trombettoni for a critical reading of the manuscript. This work is supported by the Deutsche Forschungsgemeinschaft (DFG,
  German Research Foundation) under Germany's Excellence Strategy EXC2181/1-390900948 (the
  Heidelberg STRUCTURES Excellence Cluster).
  
\appendix
\section{Defect density in the harmonic oscillator}
The dynamics of a time-dependent harmonic oscillator can be solved exactly\,\cite{lewis1967classical,lewis1969exact,lewis1968class} and any dynamical state $\psi(x,t)$ in the representation of the coordinate $x$ can be expressed as
\begin{align}
\label{dyn_exp_app}
\psi(x,t)=\sum \alpha_{n}\psi_{n}(x,t),
\end{align}
where $\alpha_{n}$ are time independent constants and the dynamical eigenstates are given by
\begin{align}
\label{Dyn_Eigen_app}
\psi_{n}(x,t)=\frac{1}{\sqrt{2^{n}n!}}\left(\frac{1}{2\pi\xi^{2}(t)}\right)^{\frac{1}{4}}e^{-\Omega(t)\frac{x^{2}}{2}}\nonumber\\
H_{n}\left(\frac{x}{\sqrt{2}\xi(t)}\right)e^{-i\left(n+\frac{1}{2}\right)\lambda(t)}.
\end{align}
The effective frequency $\Omega(t)$ can be expressed in terms of the effective width $\xi(t)$ as
\begin{align}
\Omega(t)=-i\frac{\dot{\xi}(t)}{\xi(t)}+\frac{1}{2\xi^{2}(t)},
\end{align}
and the quantity
\begin{align}
\lambda(t)=\int^{t}\frac{dt'}{2\xi^{2}(t')}
\end{align}
is the total phase accumulated at time $t$. The exact time evolution of the harmonic oscillator is then fully described by a single real function, which is the effective width $\xi(t)$ and satisfies the Ermakov-Milne equation
\begin{align}
\label{ermakov_eq_app}
\ddot{\xi}(t)+\omega(t)^{2}\xi(t)=\frac{1}{4\xi^{3}(t)}.
\end{align}
If the initial state at $t=-t_{i}$ is a pure state of the basis in Eq.\,\eqref{Dyn_Eigen_app}, say, the ground state, then all the coefficients $\alpha_{n}$ of Eq.\,\eqref{dyn_exp_app} vanish except for the coefficient $\alpha_0$. This also holds  at all later times and in the exact dynamical basis described by Eq.\,\eqref{Dyn_Eigen_app} no excited states will be generated\,\cite{dabrowski2016time}. However, at each time $t>-t_{i}$ the dynamical pure state $\psi_{0}(x,t)$ will, in general, be different from the instantaneous equilibrium ground state, since the effective width does not coincide with its instantaneous equilibrium result $\xi(t)=(2\omega(t))^{-1/2}$.
Yet, the exact time-dependent state can be decomposed into the adiabatic basis $\psi_{n}^{\rm ad}(x,t)$, whose wave functions read  
\begin{align}
\label{Eq_Eigen_app}
\psi_{n}^{\rm ad}(x,t)=\frac{1}{\sqrt{2^{n}n!}}\left(\frac{\omega(t)}{\pi}\right)^{\frac{1}{4}}e^{-\omega(t)\frac{x^{2}}{2}}H_{n}\left(x\sqrt{\omega(t)}\right).
\end{align}

Therefore, if we decompose any pure state $\psi_{n}(x,t)$ of the dynamical basis using the instantaneous equilibrium basis, the population of each adiabatic state will be finite as long as $2\xi(t)^{2}\neq \omega(t)^{-1}$. Then, assuming that the evolution started in the ground state at $t=-t_i$, the average number of excitations in the instantaneous equilibrium basis at time $t$ is given by~\cite{dabrowski2016time}
\begin{align}
\label{Exc_N_app}
n_\text{exc}(t)=\sum_{n\in 2\mathbb{N}}n|c_{n}(t)|^{2},
\end{align} 
where the coefficients
\begin{align}
\label{trans_amp_app}
c_{n}(t)=\int_{-\infty}^{+\infty}dx\,\psi_{n}^{\rm ad *}(x,t)\psi_{0}(x,t)
\end{align}
are the overlap amplitudes between the dynamical state and each instantaneous adiabatic state. In principle,
the definition in Eq.\,\eqref{Exc_N_app} can be also evaluated by choosing a different basis set for the transition amplitudes, rather than the eigenstates given in Eq.\,\eqref{Eq_Eigen_app}\,\cite{dabrowski2016time}. However, the basis of the eigenstates  in Eq.\,\eqref{Eq_Eigen_app} is the most natural choice in the context of the Kibble-Zurek mechanism.

Using the definition in Eq.\,\eqref{Exc_N_app} together with Eq.~\eqref{trans_amp_app} one can derive an explicit expression for the number of excitations $n_\text{\rm exc}(t)$. To this aim, we evaluate the transition amplitudes as
\begin{align}
&c_{n}(t)=\int_{-\infty}^{+\infty}dx\psi_{n}^{\rm ad *}(x,t)\psi_{0}(x,t)
=\frac{1}{\sqrt{2^{n}n!\pi}}\left(\frac{\omega(t)}{2\xi^{2}(t)}\right)^{\frac{1}{4}}\nonumber\\
&\int_{-\infty}^{+\infty}dx e^{-(\omega(t)+\Omega(t))\frac{x^{2}}{2}}H_{n}\left(\sqrt{\omega(t)}x\right),
\end{align}
performing a change of variable the above integral can be cast into the form
\begin{align*}
&\int_{-\infty}^{+\infty}dx e^{-(\omega(t)+\Omega(t))x^{2}}H_{n}\left(\sqrt{\omega(t)}x\right)=\nonumber\\
&(\Omega(t))^{-\frac{1}{2}}\int_{-\infty}^{+\infty} e^{-\left(\frac{\Omega(t)}{\omega(t)}+1\right)\frac{s^{2}}{2}}H_{n}\left(s\right)ds.
\end{align*}
Next, we employ the generating function for Hermite polynomials in the integral,
\begin{align}
&\int_{-\infty}^{+\infty} e^{-\left(\frac{\Omega(t)}{\omega(t)}+1\right)\frac{s^{2}}{2}}H_{n}\left(s\right)ds =\nonumber\\
&\lim_{z\to0}\frac{d^{n}}{dz^{n}}\int_{-\infty}^{+\infty} e^{-\left(\frac{\Omega(t)}{\omega(t)}+1\right)\frac{s^{2}}{2}}e^{2sz-z^{2}}ds=
\nonumber\\
&\sqrt{\frac{2\pi}{\left(\frac{\Omega(t)}{\omega(t)}+1\right)}}\lim_{z\to0}\frac{d^{n}}{dz^{n}}e^{-z^{2}\frac{\left(\Omega(t)-\omega(t)\right)}{\left(\omega(t)+\Omega(t)\right)}}=
\nonumber\\
&\begin{cases}
\displaystyle\sqrt{\frac{2\pi}{\left(\frac{\Omega(t)}{\omega(t)}+1\right)}}\frac{n!}{\frac{n}{2}!}\left(\frac{\Omega(t)-\omega(t)}{\Omega(t)+\omega(t)}\right)^{n/2} & \text{ $n\in 2\mathbb{Z}$},\\
0 & \text{ $n\in 2\mathbb{Z}+1$}.
\end{cases}
\end{align}
Thus, the probability of having $n$ excitations in the evolved state at the time $t$ is given by
\begin{align}
|c_{n0}(t)|^{2}=\frac{(n-1)!!}{n!!}\frac{\sqrt{2\omega(t)}}{\xi(t)\left|\Omega(t)+\omega(t)\right|}\left|\frac{\Omega(t)-\omega(t)}{\Omega(t)+\omega(t)}\right|^{n}.
\end{align}
Inserting this expression into Eq.~\eqref{Exc_N_app} we obtain the average number of excitations at time $t$,
\begin{align}
\label{expl_exc_num_app}
n_{\rm exc}(t)=\frac{\xi^{2}}{2\omega(t)}\left[\left(\frac{1}{2\xi^{2}}-\omega(t)\right)^{2}+\left(\frac{\dot{\xi}}{\xi}\right)^{2}\right]
\end{align}
as well as the ground state fidelity $f(t)=|c_{00}(t)|^2$, which reads
\begin{align}
\label{exp_fidelity_app}
f(t)=|c_{00}(t)|^{2}=\frac{\sqrt{2\omega(t)}}{\xi(t)}\left[\left(\frac{1}{2\xi^{2}}+\omega(t)\right)^{2}+\left(\frac{\dot{\xi}}{\xi}\right)^{2}\right]^{-1/2}.
\end{align}
Notice that a similar characterisation of the time dependent harmonic oscillator problem si shown in Ref.\,\cite{lewis1969exact}.
\section{Independence on the ramp rate}
Let us consider the time dependent frequency with the form
\begin{align}
\label{eq_dyn_freq_app}
\omega(t)=(\delta t)^{2z\nu}.
\end{align}
First of all, we shall prove that the Schr\"odinger equation of the problem
\begin{align}
i\frac{d\psi(x, t)}{dt}=\frac{1}{2}\left(p^{2}+(\delta t)^{2z\nu}x^{2}\right)\psi(x,t)
\end{align} 
can be rescaled via the transformations $x=\lambda \tilde{x}$ and $t=\mu\cdot s$ in such a way to map it to the $\delta=1$ case. According to the relations $\mu^{-1}=\lambda^{-2}$ and $\mu^{-1}=\delta^{2z\nu}\mu^{2z\nu}\lambda^{2}$, one shall choose $\mu=|\delta|^{-\frac{z\nu}{z\nu+1}}$ so that the state $\tilde{\psi}(\tilde{x},s):=\psi(x,t)$ solves the Schr\"odinger equation with $\delta=1$ and the thermodynamic quantities such as the fidelity or the average excitation number do not depend on $\delta$ in the $t\to \infty$ ($s\to \infty$) limit. For future convenience one may also introduce the Ermakov equation 
\begin{align}
\frac{d^{2}\xi(t)}{dt^{2}}+(\delta t)^{2z\nu}\xi(t)=\frac{1}{4\xi(t)^{3}}
\end{align}
and rescale it according to the transformation $t=|\delta|^{-\frac{z\nu}{z\nu+1}}s$ and $\xi(t)=\delta^{-\frac{z\nu}{2(z\nu+1)}}\tilde{\xi}(s)$ leading to
\begin{align}
\label{s_erm_eq_app}
\frac{d^{2}\tilde{\xi}(s)}{dt'^{2}}+s^{2z\nu}\tilde{\xi}(s)=\frac{1}{4\tilde{\xi}(s)^{3}}
\end{align}
as expected.
\section{Solution of the Ermakov-Equation}
\label{SM2}

Having proven that the dynamics for general $\delta$ may be reduced to the $\delta=1$ case, we can safely focus on this case to obtain the exact solution of the problem. The harmonic oscillator frequency varies as  
\begin{align}
\Omega(t)^{2}=|t|^{2z\nu}\end{align}
The Ermakov-Milne equation reads
\begin{align}
\label{ermakov_eq_app}
\ddot{\xi}(t)+\Omega(t)^{2}\xi(t)=\frac{1}{4\xi^{3}(t)},
\end{align}
 The solution of Eq.~\eqref{ermakov_eq_app} can be constructed from that of the associated classical harmonic oscillator
\begin{align}
\label{cl_h_osc_app}
\ddot{x}(t)+\Omega(t)^{2}x(t)=0.
\end{align}
Any solution to the classical harmonic oscillator Eq.\,\eqref{cl_h_osc_app} can be written in terms of the two independent solutions
\begin{align}
\label{ho_sol_app}
f_{1}(t)&=\sqrt{|t|}\,J_{\frac{1}{2+2z\nu}}\left(\frac{|t|^{1+z\nu}}{1+z\nu}\right),\qquad\\
f_{2}(t)&=\sqrt{|t|}\,J_{-\frac{1}{2+2z\nu}}\left(\frac{|t|^{1+z\nu}}{1+z\nu}\right)
\end{align}
in terms of the Bessel functions $J_{\gamma}(x)$. A proof that the functions in Eq.\,\eqref{ho_sol_app} are solutions of Eq.\,\eqref{cl_h_osc_app} can be obtained by the chain rule for the derivatives of the Bessel functions\,\cite{abramowitz1988handbook}. It is convenient to define the following quantities
\begin{align}
p&=\frac{1}{2+2z\nu},\\
\zeta&=\frac{|t|^{1+z\nu}}{1+z\nu}=2pt^{\frac{1}{2p}}.
\end{align}
Then, we introduce the generalised Airy functions
\begin{align}
x_{1}(t)=\mathrm{Ai}_{p}(-t)=p\sqrt{|t|}(J_{-p}(\zeta)+J_{p}(\zeta))\\
x_{2}(t)=\mathrm{Bi}_{p}(-t)=\sqrt{p|t|}(J_{-p}(\zeta)-J_{p}(\zeta))
\end{align}
which yield the following constant Wronskian
\begin{align}
\label{wr_rel_app}
\mathrm{Wr}\{\mathrm{Ai}_{p}(-t),\mathrm{Bi}_{p}(-t)\}=\frac{2}{\pi}\sqrt{p}\sin(p\pi).
\end{align}
It is now possible to write the solutions of Eq.~\eqref{cl_h_osc_app} as a pair of complex conjugate solutions $w$ and $w^*$ with
\begin{align}
\label{sol_def_app}
w=a x_{1}(t)+b x_{2}(t),
\end{align}
where $a\in\mathbb{R}$ and $b\in\mathbb{C}$ are constants. Accordingly, the function
\begin{align}
\label{xi_sol_app}
\xi(t)=\sqrt{ww^*}
\end{align}
is a solution of the Ermakov-Milne Eq.\,\eqref{ermakov_eq_app} if
\begin{align}
\label{wronsk_condition_app}
\mathrm{Wr}(w,w^{*})=2ia\mathrm{Im}(b)\mathrm{Wr}(x_{1},x_{2})=i,
\end{align}
which fixes one of the three coefficients in Eq.\,\eqref{sol_def_app}.

\subsection{The slow ramp to the critical point}
\label{SM3}
 According to previous section, the solution to Eq.~\eqref{ermakov_eq_app} can be constructed using Eqs.~\eqref{sol_def_app} and\,\eqref{xi_sol_app}. In addition to the relation in Eq.\,\eqref{wr_rel_app}, one needs two additional conditions in order to fix the coefficients in Eqs.\,\eqref{sol_def_app}. For a homogenous ramp starting in the ground-state at $t=-\infty$ the boundary conditions read
\begin{align}
\label{b_cond_app}
\lim_{t\to-\infty}\frac{1}{2\xi(t)^{2}}=\Omega(t),\qquad
\lim_{t\to-\infty}\dot{\xi}(t)=0,
\end{align}
consistently with the system being in the adiabatic ground state in the initial stage of the dynamics. In the $t\to-\infty$ limit, $\Omega^{2}$ diverges and one must use the asymptotic expansions of the generalised  Airy functions
\begin{align}
\lim_{t\to-\infty}x_{1}(t)&\approx \sqrt{p}\cos\left(\frac{p\pi}{2}\right)\frac{2\cos\left(\zeta-\frac{\pi}{4}\right)}{\sqrt{\pi}|t|^{z\nu/2}},\\
\lim_{t\to-\infty}x_{2}(t)&\approx -\sin\left(\frac{p\pi}{2}\right)\frac{2\sin\left(\zeta-\frac{\pi}{4}\right)}{\sqrt{\pi}|t|^{z\nu/2}}.
\end{align}
According to Eq.\,\eqref{xi_sol_app} one has that the solution to the Ermakov-Milne equation reads
\begin{align}
\label{xi_expl_app}
\xi^{2}=(a x_{1}(t)+b_{1}x_{2}(t))^{2}+b_{2}^{2}x_{2}(t)^{2}.
\end{align}
In order to satsify Eqs.\,\eqref{b_cond_app}, the oscillatory terms in the expression for $\xi$ must cancel for large $t$ and it is convenient to choose
\begin{align}
\mathrm{Re}(b)=b_{1}=0,\qquad
\end{align}
Moreover, one has to impose the condition\,\eqref{wronsk_condition_app} leading to the following coefficients,
\begin{align}
a=a_{-}&=\sqrt{\frac{\pi}{2p}}\frac{1}{2\cos\left(\frac{p\pi}{2}\right)}\\
b_{2}=b_{-}&=\sqrt{\frac{\pi}{2}}\frac{1}{2\sin\left(\frac{p\pi}{2}\right)}
\end{align}
which recover the result $a=\mathrm{Im}(b)=\sqrt{\pi/2}$ for $p=1/3\,\,(z\nu=1/2)$ in Ref.\,\cite{defenu2018dynamical}. Notice that for $p=1/3$ the function $\mathrm{Ai}_{p}$($\mathrm{Bi}_{p}$) reduces to the conventional first\,(second) type Airy function\,\cite{abramowitz1988handbook}.
The resulting expression for the scale factor is
\begin{align}
\xi(t)^{2}=a_{-}^{2}\mathrm{Ai}_{p}\left(-t\right)^{2}+b_{-}^{2}\mathrm{Bi}_{p}\left(-t\right)^{2}.
\end{align}
We can now compute the latter expression and its derivative in the $t\to0^{-}$ limit
\begin{align}
\lim_{t\to 0^{-}}&\xi^{2}(t)=\frac{\Gamma(p)\Gamma(p+1)}{2\pi p^{2p}},\\
\lim_{t\to 0^{-}}&2\dot{\xi}(t)\xi(t)=-\cot(p\pi)\mathrm{sign}(0^{-})
\end{align}
leading to a diverging defect density and vanishing fidelity according to Eqs.\,\eqref{expl_exc_num_app} and\,\eqref{exp_fidelity_app}; notice however that the excess energy remains finite.
\subsection{The full cycle}
The above section treated the case of a semi-infinite quench with frequency $\Omega(t)^{2}=t^{2z\nu}$ starting at $t=-\infty$ and terminating at $t=0$. Now, we are gonna extend such treatment to the entire interval $t\in(-\infty,+\infty)$. In order to accomplish such scope we need to extend the solution of previous section to the semi-interval $t\in(0,\infty)$. Then, we shall consider a general solution in the form of Eq.\,\eqref{sol_def_app} satisfying the boundary conditions
\begin{align}
\lim_{t\to 0^{+}}&\xi^{2}(t)=\frac{\Gamma(p)\Gamma(p+1)}{2\pi p^{2p}},\\
\lim_{t\to 0^{+}}&2\dot{\xi}(t)\xi(t)=\cot(p\pi)
\end{align}
in order to ensure continuity with the solution in the $t<0$ case. Interestingly, this result is accomplished by the coefficients choice
\begin{align}
a&=a_{+}=\sqrt{\frac{\pi}{2p}}\frac{1}{2\sin\left(\frac{p\pi}{2}\right)}\\
\mathrm{Re}(b)&=b_{1}=0\\
\mathrm{Im}(b)&=b_{2}=b_{+}=\sqrt{\frac{\pi}{2}}\frac{1}{2\cos\left(\frac{p\pi}{2}\right)}
\end{align}
which automatically satisfy the Wronskian condition in Eq.\,\eqref{wronsk_condition_app}.

The defect density in the large time limit $t\approx +\infty$ can be obtained by the asymptotic behaviour of the scale $\xi(t)$, which reads
\begin{align}
&\lim_{t\to\infty}\xi(t)^{2}\approx\frac{t^{z\nu}(1+\cos(p\pi)^{2}+2\cos(p\pi)\sin(2\zeta))}{2\sin(p\pi)^{2}}\\
&\lim_{t\to\infty}\xi(t)\dot{\xi}(t)\approx\frac{\cos(p\pi)\cos(2\zeta)}{\sin(p\pi)^{2}}.
\end{align}
Once these expressions are plugged into Eqs.\,\eqref{expl_exc_num_app} and\,\eqref{exp_fidelity_app}, one obtains the two results
\begin{align}
&\lim_{t\to\infty}n_{\rm exc}(t)=\mathrm{cot}(p\pi)^{2}\\
&\lim_{t\to\infty}f(t)=\sin(p\pi)
\end{align}
proving that the fidelity of a quantum harmonic oscillator driven across its critical point is always a constant irrespectively of the power $z\nu$ of the ramp time dependence.

\subsection{Avoided crossing}
It is interesting to consider a slightly generalised version of Eq.\,\eqref{eq_dyn_freq_app}, where the actual crossing of the quantum critical point is avoided,
\begin{align}
\label{freq_gapped_cycle_app}
\omega(t)^{2}=\left(t_{0}+\delta\,|t|\right)^{2z\nu}
\end{align}
In this case, for $t_{0}>0$ the instantaneous spectrum of the model described by the Hamiltonian in Eq.\,(1) of the main text remains non degenerate at all times. Then, according to the transformations above Eq.\,\eqref{s_erm_eq_app}, the scaled width $\tilde{\xi}(s)$ has to be evaluated at $s_{0}=\delta^{-1/(1+z\nu)}t_{0}$. Inserting $\tilde{\xi}(s_{0})$ into the defect density and fidelity Eqs.\,\eqref{expl_exc_num_app} and\,\eqref{exp_fidelity_app} yields the result for these quantities at the time $t_{0}$. In the limit $\delta\to 0$ the scaled final time $s_{0}$ diverges and, given the conditions in Eq.\,\eqref{b_cond_app}, the adiabatic result is recovered apart from the expected perturbative corrections
\begin{align}
\label{ad_corr}
\lim_{\delta \to 0}n_{\rm exc}(t_{0})= o\left(\delta^{2}\right);\qquad \lim_{\delta \to 0}f(t_{0})=1-o\left(\delta^{2}\right).
\end{align}
Accordingly, the heat (or excess energy) generated during the dynamics 
\begin{align}
Q(t)=\langle \psi_{0}| H(t)| \psi_{0}\rangle-\langle \psi^{\rm ad}_{0}| H(t)| \psi^{\rm ad}_{0}\rangle
\end{align}
vanishes with lowering $\delta$ (increasing $s_{0}$). The same behaviour is observed if the system is evolved according to the dynamical protocol in Eq.\,\eqref{freq_gapped_cycle_app} for an entire cycle $t\in[-t_{i},t_{i}]$ and, then, the $\delta\to0$ limit is performed, see Fig.\,2a of the main text.

\section{ Application to long-range $O(N)$ field theories}

The Harmonic Hamiltonian in Eq.\,(1) is  relevant to a wide range of physical applications. Several examples of this fact can be found in the literature, starting from the Bose-Hubbard model\,\cite{polkovnikov2008breakdown} across the case of scale invariant continuous Bose and Fermi gases\,\cite{haas1996hamiltonian,pitaevskii1997breathing,castin2012unitary,murthy2019quantum} and arriving to cosmological applications\,\cite{matacz1994coherent,carvalho2004scalar}.

In the present section we are going to discuss the application of the present results to study defect formation in $O(N)$ field theories with long-range interactions. The importance of $O(N)$ symmetric models in the study of critical phenomena dates back to early investigations by K.\,Wilson\,\cite{wilson1974renormalization, brezin1976renormalization}. Since then, they have been the testbed of countless field theory approaches, from real-space  and momentum space RG\,\cite{brezin1976renormalization,efrati2014real} to variational perturbation theory\,\cite{kleinert2001critical} and functional RG\,\cite{codello2013o(n), codello2015critical}. Even currently, $O(N)$ models are the subject of deep investigations as the continue to be the source of novel phenomena\,\cite{fei2014critical, yabunaka2017surprises, defenu2020fate,connelly2020universal}. On the one dimensional lattice the Hamiltonian of $O(N)$ field theories is given by
\begin{align}
\label{lr_phi4}
\mathcal{H}(\boldsymbol{\varphi})&=\sum_{i=1}^{L}\left\{\frac{\hat{\boldsymbol{\pi}}_{i}^{2}}{2}-\sum_{r=1}^{L/2-1}\frac{J_{r}}{2}\hat{\boldsymbol{\varphi}}_{i}\cdot\hat{\boldsymbol{\varphi}}_{i+r}+V(|\hat{\boldsymbol{\varphi}_{i}}|)\right\}\nonumber\\
&\mathrm{where}\,\,V(|\hat{\boldsymbol{\varphi}_{i}}|)=\frac{\mu}{2}|\hat{\boldsymbol{\varphi}}_{i}|^{2}+\frac{g}{24\,N}|\hat{\boldsymbol{\varphi}}_{i}|^{4}
\end{align}
and $\hat{\boldsymbol{\varphi}}$ and $\hat{\boldsymbol{\pi}}$ are $N$-components vector operators, whose components obey harmonic oscillator commutation relations $[\hat{\varphi}_{i\,\mu},\hat{\pi}_{b\,\nu}]=i\delta_{ij}\delta_{\mu\nu}$. The indices $i,j$ labels the sites of a 1-dimensional lattice with $L$ sites and the indices $\mu,\nu$ the field components. Apart from the traditional quartic $|\boldsymbol{\varphi}|^{4}$ potential, the fields are coupled by non-local translational invariant couplings $J_{r}\propto r^{-\alpha}$. 

In general,  the dynamical evolution of the model in Eq.\,\eqref{lr_phi4} cannot be solved exactly. However, in the limit of infinitely many field components ($N\to\infty$) the model becomes solvable, since the quartic interaction term only renormalizes the mass of the model\,\cite{moshe2003quantum}, so that the time dependent Hartee-Fock approximation properly describes the dynamical evolution\,\cite{berges2007quantum}. Thus, upon rescaling factors of $N$ in the field variables and the couplings, each component of the vector field obeys the Hamiltonian
\begin{align}
\label{lr_phi4_scalar}
\mathcal{H}_{s}(\varphi)&=\sum_{i=1}^{L}\left\{\frac{\hat{\pi}_{i}^{2}}{2}-\sum_{r=1}^{L-1}\frac{J_{r}}{2}\hat{\varphi}_{i}\cdot\hat{\varphi}_{i+r}+\frac{\mu_{\rm eff}}{2}\hat{\varphi}_{i}^{2}\right\},\\&\mathrm{with}\quad \mu_{\rm eff}(t)=\mu(t)+\frac{g}{6}\langle \hat{\varphi}_{i}^{2}\rangle.\nonumber
\end{align}
The above Hamiltonian can be rewritten in Fourier space as an ensemble of independent Harmonic oscillators with dispersion relation $\omega_{q}^{2}(t)=J(q)+\mu_{\rm eff}(t)$. Then, the system evolves according to the Ermakov equation
\begin{align}
\label{erm_eq}
\ddot{\xi}_{q}(t)+\omega_{q}(t)^{2}\xi_{q}(t)&=\frac{1}{\xi_{q}^{3}(t)}
\end{align}
where $\xi_{q}(t)$ is the effective width of the momentum space components of the field $\hat{\boldsymbol{\varphi}}_{i}$. The dynamics of quantum $O(N)$ field theories in the large $N$ limit reduces to the one of independent harmonic oscillators effectively coupled by the self-consistent mass equation
\begin{align}
\label{eff_mass}
\mu_{\rm eff}=\mu(t)+\frac{g}{6}\sum_{q} \frac{\xi_{q}^{2}}{L}.
\end{align}
The dynamical  Eq.\,\eqref{erm_eq} applies to any evolution occurring in the symmetric phase of the model, where no-order parameter is found. If the dynamics crosses the phase boundary an additional classical harmonic oscillator contribution for the order parameter dynamics has to be included in Eq.\,\eqref{erm_eq}\,\cite{chandran2013equilibration}.

The global fluctuation contribution to the effective mass in Eq.\,\eqref{eff_mass} may alter the critical behaviour of this model with respect to the harmonic case depending on the $\alpha$ value. An extensive account of the critical properties of the spherical model as a function of the decay exponent $\alpha$ and of its connection with the critical properties of $O(N)$ models at finite $N$ can be found in Refs.\,\cite{vojta1996quantum,defenu2017criticality}. 

Given Eq.\,\eqref{erm_eq}, the quantum adiabatic cycle described in the main text coincides with the dynamics of $O(N)$ field theories, whose mass evolves as 
\begin{align}
\label{mass_evolution}
\mu(t)=\mu_{c}+(\delta\,|t|)^{2z\nu}
\end{align}
where $t\in[-1/\delta,1/\delta]$ and $\delta$ is the quench rate.  The critical value of the mass $\mu_{c}$, such that $\mu_{\rm eff}=0$, separates the symmetric phase of the model ($\mu>\mu_{c}$) from the spontaneously broken one ($\mu<\mu_{c}$). Then, the dynamical protocol on the l.h.s. of Eq.\,\eqref{mass_evolution} describes a cycle of the system from the symmetric phase to the critical point and back on the symmetric phase. 

Therefore, the results derived in the main text will faithfully describe the dynamics of $O(N)$ field theories in the large-$N$ limit as long as the self-consistent contribution to the effective mass $\mu_{\rm eff}$ in Eq.\,\eqref{eff_mass} can be neglected. This is the case of long-range interactions with $\alpha<d$, where the spectrum of the system remains discrete also in the thermodynamic limit\,\cite{defenu2021metastability}. As a consequence, all excitations with $q\neq0$ remain adiabatic during the dynamics and only yield a $\delta^{2}$ contribution in Eq.\,\eqref{eff_mass} with respect to the equilibrium effective mass, which can be ignored in the $\delta\to 0$ limit. Then, for $\alpha<d$ the leading contribution to the universal dynamics is given by the zero mode $q=0$, which is decoupled by higher momentum modes in the $\delta\to0$ limit and can be described by the solution of a single harmonic mode cycled across its quantum critical point, as described in the main text. In this perspective, the $\alpha\to 0$ limit of $O(N)$ models with $z\nu=1/2$ exactly reproduces the physics described in Ref.\,\cite{defenu2018dynamical} for the Lipkin-Meshkov-Glick Hamiltonian.

It is worth noting that, while in Eq.\,\eqref{mass_evolution} we have focused on a cycle to the critical point and back, the results discussed in the main text also apply to dynamical manipulations across the quantum critical point, deep into the symmetry broken phase. Indeed the contribution from the order parameter dynamics can be ignored in the quasi-static limit ($\delta\to 0$) as long as $\alpha<d$, see the argument in Ref.\,\cite{defenu2018dynamical}.

\bibliographystyle{apsrev_titles}
\bibliography{bibliography}

\end{document}